\documentclass[12pt]{article}

\newcommand{\tit}[1]{\begin{center}
\Large \bf #1
\end{center}
\vskip 0.7cm }
\newcommand{\autaf}[2]{\begin{center}
#1\vskip 0.5cm #2
\end{center}}
\newcommand{\bod}[4]{
\flushright{Presentation:#1}
\flushright{Topic:#2}
\vskip 1cm
\begin{minipage}[c]{15.00cm} #3
\end{minipage}
\begin{thebibliography}{99}
#4 \end{thebibliography}}

\newcommand{\art}[6]{\bibitem{#1} #2, {\it #3} {\bf #4} (#5) #6}

\newenvironment{abs}
{\pagestyle{empty}}
{\newpage}
\usepackage{latexsym}

\usepackage{graphicx}
\usepackage{dcolumn}
\usepackage{bm}

\begin{document}

\begin{abs}
\tit{ Critical exponents for the long-range Ising chain using a
transfer matrix approach}
\autaf{\underline{R. F. S. Andrade} and S. T. R. Pinho }

\autaf{ Instituto de F\'{i}sica, Universidade Federal da Bahia\\
Campus Universit\'ario de Ondina - Salvador, 40210-340, Brazil,
randrade@ufba.br }

%
\bod { Oral}
 { long range Ising model, critical
behavior}
 { The critical behavior of the Ising chain with
long-range ferromagnetic interactions decaying with distance
$r^\alpha$, $1<\alpha<2$, is investigated using a numerically
efficient transfer matrix (TM) method. Finite size approximations
to the infinite chain are considered, in which both the number of
spins and the number of interaction constants can be independently
increased. Systems with interactions between spins up to $18$
sites apart and up to 2500 spins in the chain are considered. We
obtain data for the critical exponents $\nu$ associated with the
correlation length based on the Finite Range Scaling (FRS)
hypothesis. FRS expressions require the evaluation of derivatives
of the thermodynamical properties, which are obtained with the
help of analytical recurrence expressions obtained within the TM
framework. The Van den Broeck extrapolation procedure is applied
in order to estimate the convergence of the exponents. The TM
procedure reduces the dimension of the matrices and circumvents
several numerical matrix operations. }

\end{abs}

\section{Introduction}

It has been well-known that the thermodynamical properties and
critical behavior of physical models is affected by the presence
long-range interactions. They define completely new classes of
models, where cooperative effects are enhanced and change the
thermodynamical properties in comparison to those of the short
range interaction models. Due to the presence of simultaneous
coupling among many degrees of freedom, the long range models
offer very difficult technical problems that makes it impossible
to derive exact expressions for the thermodynamical properties
within the equilibrium statistical ensemble formulation. This is
observed already for the most simple long-range Ising chain, which
has been investigated for many decades. In its most simple
version, each spin $\sigma_i$ interacts with all other spins on
the chain mediated by coupling constants $J_r=J/r^\alpha$, where
$r$ is the distance between the interacting spins measured in
integer number of lattice spacings.

Although no closed form solution for this model is available,
there are several rigorous results on the existence of distinct
thermodynamical phases, what depends on the range of values of
$\alpha$. The most important features of these rigorous results
are: for $\alpha>2$, the system shows only a disordered phase,
$\forall T$ \cite{Aizeman1,Aizeman2}; for $1<\alpha\leq 2$, there
is a phase transition at finite temperature $T_c$
\cite{Dyson,Frolich}; critical mean field behavior occurs for $1<
\alpha\leq 1.5$ \cite{Aizeman1,Aizeman2}; for $\alpha<1$, only one
ordered phase exists, $\forall T$. Regarding the evaluation of
approximate results, both renormalization group schemes
\cite{Cardy}-\cite{Glumac0} and numerical calculations of finite
size systems \cite{Glumac}-\cite{Monroe} have been used to
estimate the thermodynamical properties, the critical temperature
and the critical exponents when $1<\alpha\leq 2$.

More recently, the scaling behavior of this model, specially in
the range $\alpha\leq 0<1$, where the energy is not an extensive
quantity, has also been addressed. Investigations have been
motivated by an universal scaling scheme proposed by Tsallis
\cite{Tsallis1}-\cite{Tsallis3}, that should be valid for both
extensive and non-extensive long range models, as those
constituted by spins \cite{Cannas0}-\cite{Cannas3} rotors
\cite{Luciano, Pancho}, and so on.

In this work we use a numerically efficient transfer matrix (TM)
approach \cite{Andrade1} to analyse the critical behavior of the
system, estimating the critical exponents $\nu$ associated with
the correlation length in the range for $1<\alpha\leq 2$, where
the system undergo a phase transition. In a previous paper
\cite{Andrade3}, we have introduced this approach to check the
validity of Tsallis' scaling conjecture to the Ising long range
chain, and to find estimates for the critical temperature. We
obtained results that show very good accordance with other
numerical estimates, indicating that the proposed approach is
quite reliable. It takes into account the long range interactions
between spins up to a certain distance $g$ apart in the evaluation
of the thermodynamical properties of the system. Also, this method
allows to independently increase the number of spins $N$ in the
chain, so that $N \geq g+1$.

The TM approach used herein makes use of very compact matrices, so
that the configuration energies and Boltzmann weights, that are
numerically evaluated, can be stored and operated in a very
efficient way. For instance, we note that the TM procedure avoids
the numerical evaluation of the TM eigenvalues. For the evaluation
of $\nu$, this framework is quite useful, as it leads to
analytical expressions for the derivatives of the Boltzmann
weights that can be similarly stored in very compact matrices. So,
the derivatives of thermodynamical properties can be directly
computed, avoiding the use of numerical differentiation
\cite{Glumac}.

This work is organized as follows: in Section II we discuss the
essential aspects of transfer matrix method (TM) used to evaluate
the critical exponents. In Section III we present the critical
temperature, the critical exponents $\nu$ and $\beta$ associated
with the correlation length and the magnetization for values of
$\alpha$ in the following range: $1 \le \alpha \le 2$. In Section
IV we discuss the results and compare with some previous results
\cite{Glumac}. Section V closes the work with our final remarks
and perspectives.

\section{Transfer matrix framework}

The long range Ising chain is described by the Hamiltonian
\begin{equation}\label{eqa}
\mathcal{H}=-\sum_{i=0}^{\infty}\sum_{r=i+1}^{\infty}J_r\sigma_i\sigma_{i+r}
-h\sum_{i=0}^{\infty}\sigma_i,
\end{equation}
\noindent

\begin{figure}
\begin{center}
\includegraphics[width=0.25\textwidth,angle=270]{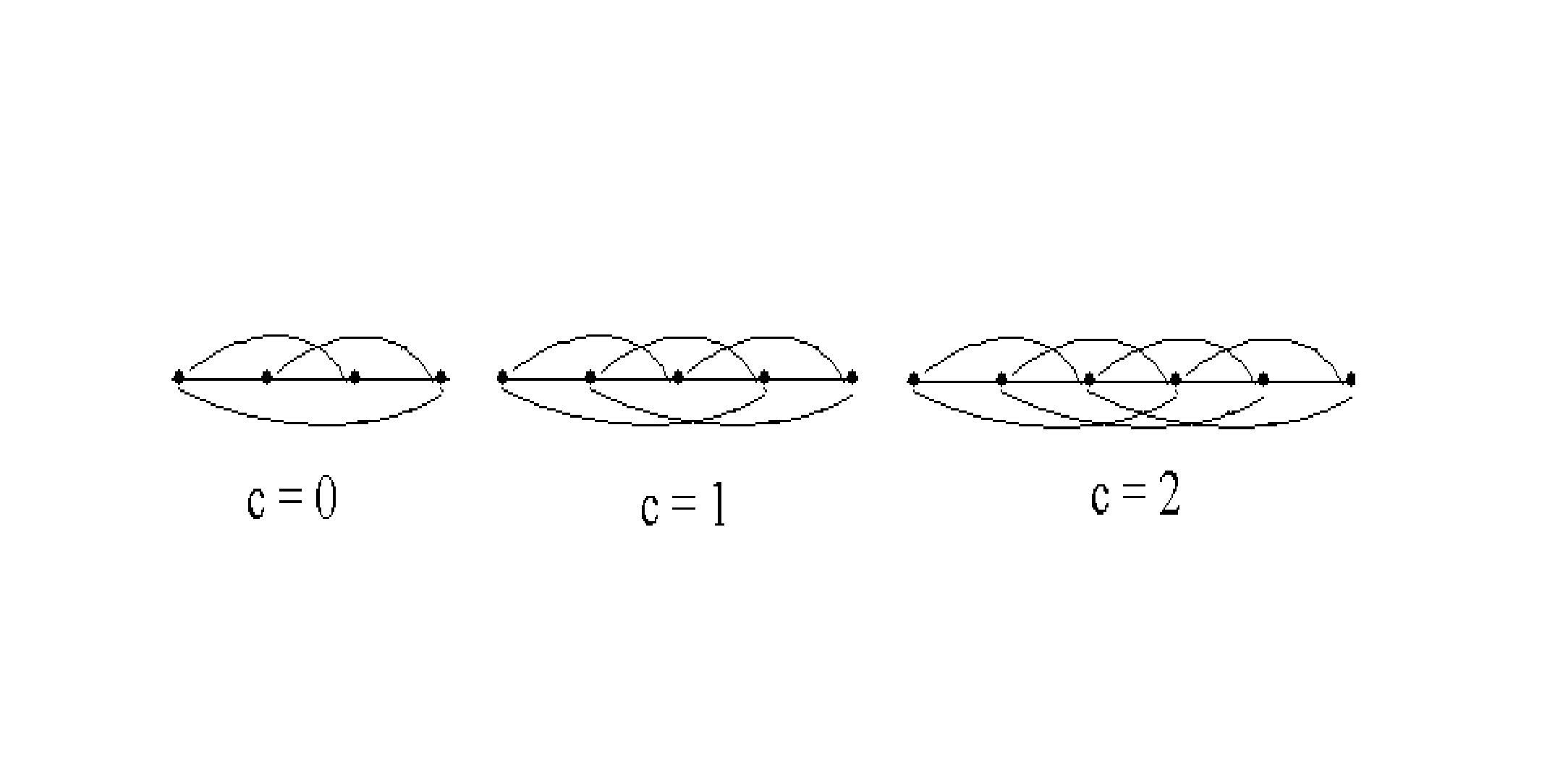}
\end{center}
\caption{\label{Figure 1} A schematic representation of the
construction of the infinite $3$-chain ($g=3$) for $c=0, 1$ and
$2$.}
\end{figure}

One can construct this model by starting with an infinite chain
but a finite number interaction constants. Then, in a series of
steps $g$, all $J_g$ are introduced at once, among all pairs of
spins that are $g$ sites apart. It is also possible to construct
the model by taking a finite chain with $g$ distinct coupling
constants and $N = g+c+1 \geq g+1$ spins. Thus, for a generic
value $g$, if $c$ new spins are included in the chain, $g\times c$
new interactions, mediated by $J_r$, $r = 1, 2, ... g$, are added,
linking every new spin to all spins up to $g$ sites apart that
have been introduced at the previous generations. In Figure 1a, 1b
and 1c, we show, for example, for fixed $g=3$, the interactions
for the first 3 values of $c$. If $g\rightarrow\infty$,
$N\rightarrow\infty$, so that this procedures leads to the same
the actual system. However, for finite values of $g$, the first
construction procedure can be recovered by letting
$c\rightarrow\infty$.

In this work, we consider the second way of construction quoted
above, which is schematically  represented in Figure 1, for which
the finite chain has $N=g+c+1$ spins that interact with all spins
$g$ sites apart. This way, it is possible to obtain a partition
function based on $2\times2$ TM $M_{g,c}$, instead of
$2^g\times2^g$ TM $\widehat{M_g}$ introduced in reference
\cite{Glumac}. Besides working with smaller TM's, this approach
has the  advantage of avoiding the use of numerical
diagonalization procedures to evaluate the largest eigenvalues.
According to this TM scheme, which has been described with enough
details in \cite{Andrade3}, the TM $2\times2$ TM $M_{g,c}$ are
given by:

\begin{equation}\label{eq9}
M_{g,c}=(\prod_{k=1}^{g}P_k)(Q_g \cdot P_g)^c\cdot
L_g\equiv.R_{g,c}L_g.
\end{equation}
\noindent \noindent where $L_g$ is a $2^{g+1}\times2$ matrix whose
elements $(L_g)_{i,j} = 1$, for $i+j$ even, and $(L_g)_{i,j} = 0$,
for $i+j$ odd; $P_k$ are recursively expressed by:
\begin{equation}\label{eq5}
(P_k)_{i,j}=\left \{ \begin{array} {lll}
(P_{k-1})_{i,j}$ $a_k^{(-1)^{j-1}} & $ for $ i\leq 2 ^{k-1} ,\\
  & $ and $ j\leq 2^k \\
(P_{k})_{2^k-i+1,2^{k+1}-j+1} & $ for $ 2 ^{k-1}\leq i \leq 2 ^k \\
  & $ and $ 2^k\leq j \leq 2^{k+1} \\
0 & $ otherwise $
\end{array}
\right..
\end{equation}
\noindent where $a_k=exp(J_k/T)$; $Q_g$ is a $2^{g+1}\times 2^g$
matrix defined by

\begin{equation}
(Q_g)_{i,j}=\left \{
\begin{array} {ll}
1 & $ for $ i=j $ or $ i=j+2^g \\
0 & $ otherwise $
\end{array}
\right. ,
\end{equation}

Therefore the free energy per spin, $f_{g,c}=-T\ln (Z_{g,c})/N$
follows from the partition function
\begin{equation}\label{eq9a}
Z_{g,c}=2\lambda_{g,c}^{+}=2((M_{g,c})_{1,1}+(M_{g,c})_{1,2})=\sum_{i,j}(R_{g,c})_{i,j}.
\end{equation}

The correlation function between the first and the $r$-th spins
along the chain, restricted to the case $c=0$, defined by a
$g$-dependent correlation function
$C_g(r;T)=\langle\sigma_1\sigma_{r}\rangle_g,r=1,...,g$, is given
by
\begin{equation}\label{eq10}
C_g(r;T)\equiv\frac{1}{Z_g}\sum_{i,j}\left[\underline{P}_1(\prod_{k=2}^{g}P_k)
L_{g,r}\right]_{i,j},
\end{equation}
where
\begin{equation}\label{eq10b}
\underline{P}_1=\left( \begin{array} {llll}
a_1 & b_1 & 0 & 0 \\
0 & 0 & -b_1 & -a_1
\end{array}
\right),\hspace{0.5cm}(L_{g,r})_{i,j}={-1}^{q_{g,r}(j)},
\hspace{0.5cm}{q_{g,r}(j)}=L\left[\frac{j-1}{2^{g-r}}\right],
\end{equation}
with $L[x]\equiv$ largest integer in $x$.

The definition of $C_g(r;T)$ can be extended to
$C_{g,c}(i,i+r;T)$, where $c>0$, $r>g$. If we use (\ref{eq10}) for
$r=g$, we have $C_g(g;T)=\lambda_{g}^-/\lambda_{g}^+$. Note
further that the correlation length for a chain composed of
patches described by the matrix $M_{g,c=0}$ is given by
\begin{equation}\label{eq12}
\xi_g =
\frac{g}{\ln(\lambda_{g}^+/\lambda_{g}^-)}=-\frac{g}{\ln(C_g(g;T_{c,g}))}.
\end{equation}

Based on the above expressions for thermodynamical properties, we
obtain analytical expressions for $\lambda_{g}^+$ and
$\lambda_{g}^-$ given by:

\begin{equation}\label{lambda+-}
\lambda^{\pm}_g = \sum_{j=1}^{2^{g-1}} [a(j,1) \pm a(2^g+1-j,1)]
\gamma^{\pm}_g(j)],
\end{equation}

\noindent where

\begin{equation}
\left \{
\begin{array}{lll}
\gamma^{\pm}_{g+1}& = & a(2j-1,2) \gamma^{\pm}_g(2j-1)+a(2j,2)
\gamma^{\pm}_g(2j)\;\;\;\; j = 1,2,\ldots, 2^{g-2} \\
\gamma^{\pm}_{g+1} & = & \pm a(2j-1,2) \gamma^{\pm}_g(2^g+2-2j)
\pm a(2j,2) \gamma^{\pm}_n(2^g+1-2j),\;\; j=2^{g-2}+1 \ldots
2^{g-1}
\end{array}
\right. ,
\end{equation}
with $\gamma^{\pm}_c(0)=0$.

Since the eigenvalues $\lambda^{\pm}_g$ are given by
(\ref{lambda+-}), we can easily derive expressions for their
derivatives with respect to $T$ and $h$, from which the
thermodynamic functions, expressed in terms of the derivatives of
the free energy, can be easily evaluated without resorting to
numerical differentiation.

To obtain the scaling properties of $\xi$ we will make use of the
FRS framework \cite{Glumac}. This scheme proposes a scaling
hypothesis which is formally similar to the well known finite size
scaling, which compares the behaviors of finite size systems with
different number of components. In FRS, one assumes similar
relations among systems with distinct number of coupling
constants, hence of different interaction ranges. Starting from
the assumption that, close to the critical temperature
$\overline{T}$, any thermodynamical function $y(t)$ of the reduced
temperature $t=(T-\overline{T}/\overline{T})$ is described by a
power law

\begin{equation}\label{eqglumac2}
y(t) = A_0t^{-\rho},
\end{equation}
and that a finite number $g$ of coupling constants modifies the
actual criticality by a correction factor $f$, it is proposed that

\begin{equation}\label{eqglumac3}
y_g(t) = y_{\infty}(t) f(g/\xi_{\infty}).
\end{equation}
With (\ref{eqglumac2}) and (\ref{eqglumac3}) it is possible to
show that, close to $\overline{T}$, the condition
\begin{equation}\label{eq13}
\frac{\xi_g(t)}{\xi_{g+1}(t')} = \frac{g}{g+1},
\end{equation}
holds. Or, equivalently, $t$ and $t'$ are related by
\begin{equation}\label{eq13a}
t' = [\frac{g}{g+1}]^{\frac{1}{\nu}}t.
\end{equation}
The critical temperature $\overline{T}_g$ at order $g$ is obtained
by the condition
\begin{equation}\label{eq13aa}
\frac{\xi_g(\overline{t}_g)}{\xi_{g+1}(\overline{t}_g)}=\frac{g}{g+1},
\end{equation}
and it is expected that the series of values $\overline{t}_g$
converges to the actual $\overline{t}=0$ in the limit
$g\rightarrow\infty$.

Linearizing and expanding around $\overline{T}_g$, leads to
\begin{equation}\label{eq13b}
\frac{\xi_{g+1}(t)}{\xi_g(t')} =
\frac{\xi_{g+1}(\overline{t}_g)+(d\xi_{g+1}(\overline{t}_g)/dt)
t}{\xi_{g}(\overline{t}_g)+(d\xi_{g}(\overline{t}_g)/dt) t'}.
\end{equation}
Combining (\ref{eq13a}) with (\ref{eq13b}), we obtain estimates of
the critical exponent $\nu_g$ as
\begin{equation}\label{eq13c}
\frac{1}{\nu_g} =
\frac{\ln(\frac{d\xi_{g}(\overline{t}_g)/dt}{d\xi_{g+1}(\overline{t}_g)/dt})-1}{\ln(\frac{g}{g+1})}.
\end{equation}

\begin{figure}
\begin{center}
\includegraphics[width=0.5\textwidth,angle=270]{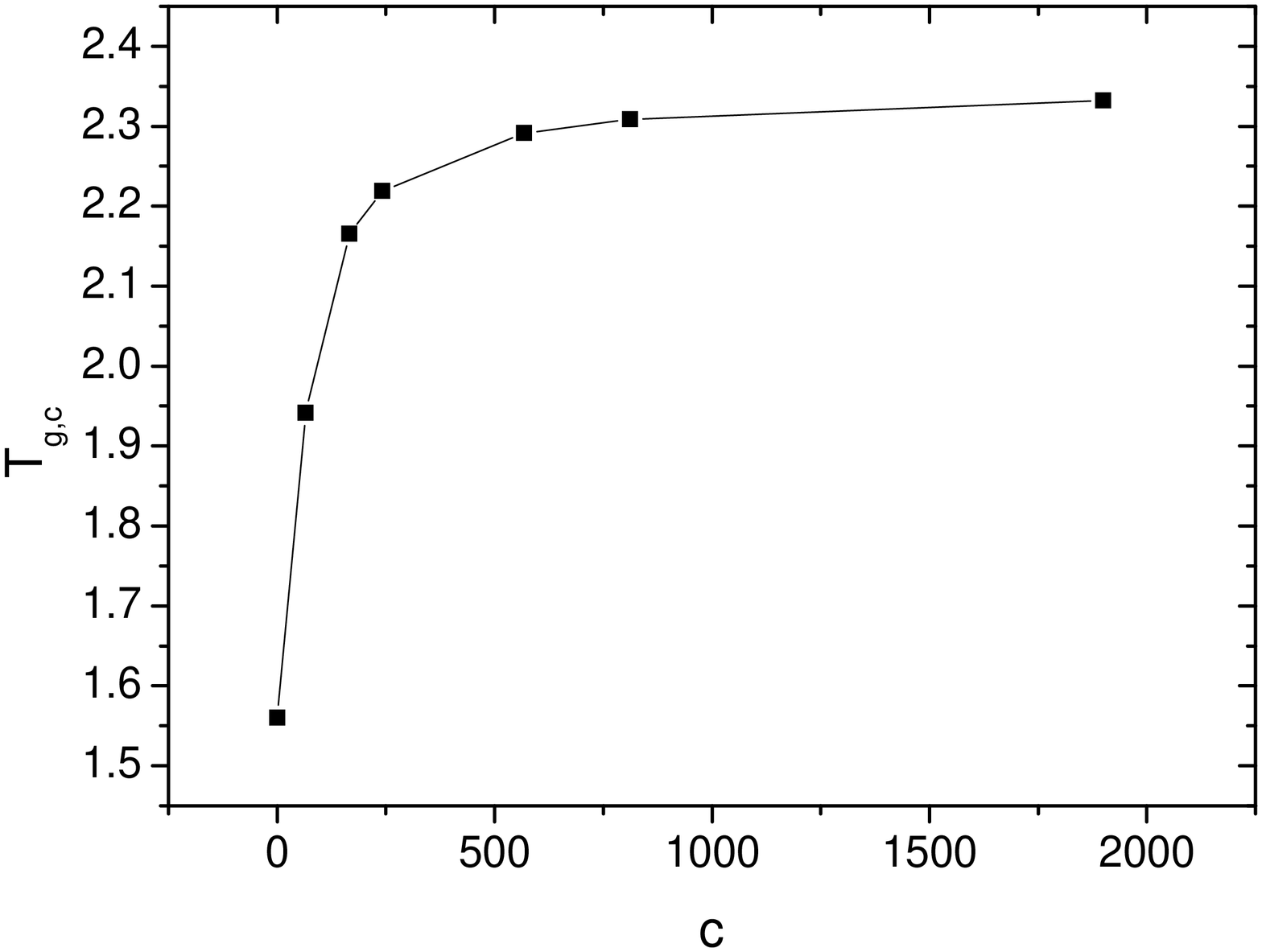}
\end{center}
\caption{\label{Figure 2} Dependence of $\overline{T_{g,c}}$ on
$c$, for $\alpha = 1.8$ and a fixed value of $g$, for example,
$g=10$.}
\end{figure}

\section{Results}

In reference \cite{Andrade3}, we have obtained a series of
estimates $\overline{T}$ for the  critical temperature
$\overline{T}$, based on the observed behavior for $C_g(r;T)$,
which corresponds exactly to the condition expressed by
(\ref{eq13aa}). In such case, we have considered $N=g+1$, i.e.,
$c=0$. The actual value of $\overline{T}$ has been evaluated by an
extrapolation procedure, with very good accuracy, although the
individual values $\overline{T_g}$ are not so close to
$\overline{T}$.

To address the question of the critical behavior near critical
temperature, we need to be in a much closer neighborhood of
$\overline{T}$. So we found it becomes necessary to consider a
larger number of spins in the chain, what requires larger values
of $g$ and $c$. Increasing $g$ leads to exponential growth of
storage capacity and computing time. We found, however, that
taking a better approximations to the infinite chain, what amounts
to increasing $c$ for a fixed value of $g$, makes it possible to
reach a closer neighborhood of $\overline{T}$.

To this purpose, we evaluate a series of critical temperatures
$\overline{T_{g,c}}$ depending on both $g$ and $c$, from which we
obtain first
$\overline{T_{g}}=\lim_{c\rightarrow\infty}\overline{T_{g,c}}$ and
$\overline{T}=\lim_{g\rightarrow\infty}\overline{T_{g}}$. Each
$\overline{T_{g,c}}$ is obtained by imposing the condition
(\ref{eq13b}), where each $\xi_g$ actually indicates $\xi_{g,c}$,
which is evaluated for corresponding values of $g$ and $c$ with
the help of ($\ref{eq9}$). In Figure 2, we show how
$\overline{T_{g,c}}$ depends on$c$, for a fixed value of $g$ when
$\alpha=1.8$; the behavior does not change for other values of
$g$.

In our numerical evaluation, $\overline{T_{g}}$ has been obtained
with a given accuracy, as we increase the value of $c$ until $\mid
\overline{T_{g,c+1}}-\overline{T_{g,c}}\mid<\epsilon$, where the
highest accuracy corresponds to $\epsilon\simeq 10^{-4}$. For such
situation, $c$ can be as large as 2500, depending on the values of
$g$ and $\alpha$. The largest value of $g = 18$ has been obtained
for $\alpha=1.6$.

In Tables 1, 2, 3 and 4, we show the values of $\overline{T_{g}}$
and $\nu_{g}$, for distinct values of $g$ and $1< \alpha <2$:
$\alpha=1.2, 1.4, 1.6$ and $1.8$, for a fixed value of accuracy
$\epsilon=10^{-4}$. In order to analyse the effect of the accuracy
$\epsilon$, for values of $\alpha$, inside and outside the mean
field region, $\alpha=1.2$ (Table 1) and $1.8$ (Table 4)
respectively, we include, in these tables, the data for another
value of $\epsilon$, $\epsilon=10^{-3}$. It is easy to see that
the value of $c$ is very sensible to the accuracy of the
corresponding values of $c$ for distinct values of $\epsilon$; in
Figure 2 we plot the critical temperature for increasing values of
$c$, corresponding to, respectively, decreasing values of
$\epsilon$.

\begin{table}
  \begin{tabular}{|c|c|c|c|c|c|c|} \hline
 $\alpha=1.2$  & & {$\epsilon=10^{-4}$} & & &
  {$\epsilon=10^{-3}$}  & \\ \hline \cline{2-5}
  $g$ & $c(\epsilon)$ & $\overline{T_{g,c}}$ & $\nu_{g,c}$ & $c(\epsilon)$ & $\overline{T_{g,c}}$ & $\nu_{g,c}$
   \\ \hline
   3 & 684 & 5.5419339  & 2.4783289 & 208 & 5.3921944 &
    2.4336811 \\ \hline
    5 & 1058 & 6.6798492 & 2.7505706 & 314 & 6.4445671 & 2.6616931
    \\ \hline
    7 & 1396 & 7.3835491 & 2.9643628 & 410 & 7.0728148 & 2.8315852
    \\ \hline
    9 & 1706 & 7.8632795 & 3.1338069 & 492 & 7.4796335 & 2.9552605
    \\ \hline
   11 & 1990 & 8.2103090  & 3.2697569 & 566 & 7.7599961 &
   3.0455192 \\ \hline
   13 & 2260 & 8.4726154 & 3.3805739 & 632 & 7.9590277 & 3.1099467 \\ \hline
   15 & 2500 & 8.6753271 & 3.4710018 & 692 & 8.1041143 & 3.1550949 \\ \hline

VBS extrapolation & - & 9.482310 & 3.803537 & - & 8.910613 & 3.288895\\
\hline

\end{tabular}

\caption{\label{Table 1a} Values of $\overline{T_{g}}$ and
$\nu_{g}$, for odd values of $g$ and $\alpha =1.2$. Data are
displayed for accuracy $\epsilon=10^{-4}$ and $\epsilon=10^{-3}$
that was required for the limiting value of $\overline{T_{g}}$.
The corresponding smallest value of $c$ at which the accuracy was
reached is also indicated.}
\end{table}

\vspace{1.5cm}

We observe that the estimates $\overline{T_{g,c}}$, for any value
of $\alpha$, are much larger than those obtained previously for
$c=0$ that we have obtained in\cite{Andrade3}. We observe that,
despite the estimates $\overline{T_{g,c}}$ are much larger than
$\overline{T_{g,c=0}}$, some of the extrapolated values are
smaller than those that were predicted from the series
$\overline{T_{g,c=0}}$ if the value of $\epsilon$ is not small
enough.

\begin{table}
  \begin{tabular}{|c|c|c|c|} \hline
 $\alpha=1.4$ & & {$\epsilon=10^{-4}$} & \\ \hline \cline{2-3}
  $g$ & $c(\epsilon)$ & $\overline{T_{g,c}}$ & $\nu_{g,c}$
   \\ \hline
    3  & 492 & 3.8408632 & 2.1731057  \\ \hline
    5  & 694 & 4.3296422 & 2.2598101  \\ \hline
    7  & 856 & 4.5894104 & 2.31620273  \\ \hline
    9  & 994 & 4.7474461 & 2.35392944  \\ \hline
   11  & 1112 & 4.8515617 & 2.37937145  \\ \hline
   13  & 1216 & 4.9241076 & 2.39645998  \\ \hline
   15  & 1310 & 4.9767888 & 2.40768914  \\ \hline
    VBS extrapolation & - & 5.022703 & 2.437089 \\
\hline

\end{tabular}

\caption{\label{Table 1b} Values of $\overline{T_{g}}$ and
$\nu_{g}$, for odd values of $g$ and $\alpha =1.4$. Data are
displayed for accuracy $\epsilon=10^{-4}$  that was required for
the limiting value of $\overline{T_{g}}$ and the corresponding
smallest value of $c$ is also indicated.}
\end{table}

\vspace{1cm}

We see also that the values of $c$ for a fixed required accuracy
increases with $g$, using fixed $\alpha$, and decreases with
$\alpha$ for a fixed $g$. As expected, in any situation, $c$
increases as $\epsilon$ decreases. Regarding the values of
$\overline{T_{g,c}}$, we see that they form a monotonic series,
increasing with respect to both $g$ and $c$, approaching
$\overline{T}$ from the lower side. In order to extrapolate the
finite series results for a limit value $\overline{T}$ and $\nu$,
we have used the Vanden Broeck and Schwartz (VBS) extrapolation
procedure used both in (\cite{Glumac}) and (\cite{Andrade3}). We
observe that extrapolated values for
$\nu=\lim_{g\rightarrow\infty}\nu_g$, which were estimated on the
for $\overline{T_{g,c}}$, get better for small $\epsilon$ is
reduced and $c$ increases.

\begin{table}
  \begin{tabular}{|c|c|c|c|} \hline
 $\alpha=1.6$ & & {$\epsilon=10^{-4}$} & \\ \hline \cline{2-3}
  $g$ & $c(\epsilon)$ & $\overline{T_{g,c}}$ & $\nu_{g,c}$
   \\ \hline
   3 & 382 & 2.8178502 & 2.0697087 \\ \hline
    5 & 510 & 3.0448535 & 2.0947359 \\ \hline
    7 & 608 & 3.1555507 & 2.1052548 \\ \hline
    9 & 688 & 3.2190642 & 2.1091374 \\ \hline
   11 & 756 & 3.2591533 & 2.1095755 \\ \hline
   13 & 814 & 3.2859512 & 2.1080216 \\ \hline
   15 & 866 & 3.3047686 & 2.1053158 \\ \hline
   17 & 910 & 3.3180850 &     - \\ \hline
   18 & 932 & 3.3234981 &     - \\ \hline
    VBS extrapolation & - & 3.328188  &  2.109820 \\
\hline

\end{tabular}

\caption{\label{Table 1c} Values of $\overline{T_{g}}$ and
$\nu_{g}$, for odd values of $g$ and $\alpha =1.4$. Data are
displayed for accuracy $\epsilon=10^{-4}$  that was required for
the limiting value of $\overline{T_{g}}$ and the corresponding
smallest value of $c$ is also indicated.}
\end{table}

\vspace{1cm}

\begin{table}
  \begin{tabular}{|c|c|c|c|c|c|c|} \hline
  $\alpha=1.8$ & & {$\epsilon=10^{-4}$} & & &
  {$\epsilon=10^{-3}$}  & \\ \hline \cline{2-3}
  $g$ &  {$c(\epsilon)$} & $\overline{T_{g,c}}$ & $\nu_{g,c}$ &
   {$c(\epsilon)$} & $\overline{T_{g,c}}$ & $\nu_{g,c}$
   \\ \hline
    3 & 314 & 2.1323856 & 2.1231430 & 94 & 2.0633542 & 2.1560530 \\ \hline
    5 & 408 & 2.2243788 &  2.1639020 & 122 & 2.1346603 & 2.2165373 \\ \hline
    7 & 482 & 2.2642364 & 2.1873097 & 142 & 2.1575740 & 2.2647878 \\ \hline
    9 & 540 & 2.2847804 & 2.2036550 & 158 & 2.1642335 & 2.3109579 \\ \hline
   11 & 592 & 2.2967828 & 2.2163352 & 172 & 2.1644328 & 2.3583671 \\ \hline
   13 & 636 & 2.3039057 & 2.2273095 & 186 & 2.1629377 & 2.4063897 \\ \hline
   15 & 676 & 2.3083424 & 2.2372586 & 198 & 2.1588773 & 2.4598516 \\ \hline
    VBS extrapolation & -  & 2.315334 & 2.246300 & - & 2.167883 & 2.910225\\
\hline

\end{tabular}

\caption{\label{Table 1d} Values of $\overline{T_{g}}$ and
$\nu_{g}$, for odd values of $g$ and $\alpha =1.8$. Data are
displayed for accuracy $\epsilon=10^{-4}$ and $\epsilon=10^{-3}$
that was required for the limiting value of $\overline{T_{g}}$ and
the corresponding smallest value of $c$ is also indicated.}
\end{table}

In Table 5, we show the VBS extrapolated values of $\nu$ for the
same values of $\alpha$ and a fixed value of $\epsilon$ that is
small enough. Here we are lead to the most interesting result of
this paper.  For the purpose of comparison, we use the known exact
values for $\nu=1/(\alpha-1)$ when $1\leq \alpha \leq 1.5$, when
the values obtained from the mean field analysis should prevail
also for those from an exact solution. It is clear that our VBS
extrapolation leads to a better agreement in comparison to the
corresponding VBS extrapolated results in (\cite{Glumac}). Note
that we use larger values of $N$ ($N \approx 15$ for all values of
$\alpha$) than those ones used in reference \cite{Glumac}, which
should be very important for the estimation of the the critical
exponents. In this sense, the
 efficient TM method that we have applied is very useful to reduce the
CPU time of the numerical calculations.

The value of $\nu_g$ for $c=0$ are indeed far away from the exact
ones. This indicates that, although the value of $\overline{T}$
can be evaluated with good precision from the $c=0$ finite size
data, the critical properties require indeed to probe the system
in a close neighborhood of $\overline{T}$. So, it is amazing to
observe that our TM approach was able to be adapted to perform
this most sensitive task.

\begin{table}

\begin{center}

  \begin{tabular}{|c|c|c|c|} \hline
 $\alpha$ & $\nu$ & $\nu_{gu}$ & $\nu_{ex}$ \\ \hline
   1.2   & 3.803537  & 7.0 & 5.0 \\ \hline
   1.4   & 2.437089  & 2.7 & 2.5 \\ \hline
   1.6   &  2.109820 & 2.15 & -  \\ \hline
   1.8   & 2.246300  & 2.22  & - \\ \hline
\end{tabular}

\end{center}

\caption{\label{Table 2} Comparative table of VBS extrapolated values of $\overline{T}$
and $\nu$, for distinct values of $1< \alpha <2$,  using
$\epsilon=10^{-4}$, obtained from the data in Table 1. We also
include the corresponding value $\nu$, calculated by Glumac and Uzelac \cite{Glumac},
$\nu_{gu}$, and the exact value of $\nu$, $\nu_{ex}$ for the mean field
values of $\alpha$.}
\end{table}

\section{Conclusions}

In this work we presented results for the critical exponent $\nu$
for the long range Ising chain based on the numerical evaluation
of the eigenvalues of a $2\times 2$ TM that condenses the
information regarding the Boltzmann weights for all configurations
of a finite size chain of $g+c+1$ spins including interaction
among spins up to $g$ sites apart. The adopted approach requires
the minimum possible storage space and avoids the necessity of
eigenvalue evaluation. The results were obtained with a double
precision Fortran code implemented on a common desk computer.

The comparison of our estimates with similar results reported by
other authors shows that they are of the same quality or better
than those obtained by a TM procedures that requires much larger
matrices. In particular, we have used the known exact values of
$\nu$ in the range $1 < \alpha \leq 1.5$ to check the validity of
our results.

As compared to our previous implementation of our TM framework, we
have shown that it is indeed reliable and that it could be
successfully extended to much larger values of $c$, what was
required in order to probe the close neighborhood of
$\overline{T}$. Despite the fact that analytical expression for
the derivatives of the eigenvalues represents an important
achievement of our approach and avoids the numerical
differentiation of $\xi$, we have observed that numerical
overflows in our Fortran code did not allow pushing the series of
$\nu_g$ to the same larger values of $g=24$ that were reached in
our first evaluation of the critical temperature when $c=0$.
Efforts to sidestep this effect and to include the evaluation of
the magnetization exponent $\beta$ are currently being undertaken
to present a more complete analysis of this model.

\vspace{0.1in}

Acknowledgements: {This work has been partially supported by the
Brazilian agencies CNPq and FAPESB}



\end{document}